\documentclass[a4paper,11pt]{article}
\usepackage{jcappub} % for details on the use of the package, please see the JINST-author-manual
\usepackage{aas_macros} % names for the journals in the library
\usepackage{lineno}
\usepackage{nicefrac}
\newcommand{\diff}{\mathrm{d}}
\newcommand{\etal}{\textit{et al.\ }}

\newcommand{\change}[1]{#1}

\arxivnumber{2501.16967} % Only if you have one
\title{Uncertainties in astrophysical gamma-ray and neutrino fluxes from proton-proton cross-sections in the GeV to PeV range}

\author[a,b]{Julien Dörner}
\author[b,c]{Leonel Morejon}
\author[b,c]{Karl-Heinz Kampert}
\author[a,b,d]{Julia Becker Tjus}

\affiliation[a]{Theoretische Physik IV, Fakult\"at f\"ur Physik \& Astronomie, Ruhr-Universit\"at Bochum, 44780 Bochum, Germany}
\affiliation[b]{Ruhr Astroparticle and Plasma Physics Center (RAPP Center), Germany}
\affiliation[c]{Bergische Universit\"at Wuppertal, Department of Physics, Gau\ss stra\ss e 20, 42103 Wuppertal, Germany}
\affiliation[d]{Department of Space, Earth and Environment, Chalmers University of Technology, 412 96 Gothenburg, Sweden}

\emailAdd{jdo@tp4.rub.de}

\abstract{
The identification of Cosmic Ray (CR) sources represents one of the biggest and long-standing questions in astrophysics. Direct measurements of cosmic rays cannot provide directional information due to their deflection in (extra)galactic magnetic fields. Cosmic-ray interactions at the sources lead to the production of high-energy gamma rays and neutrinos, which, combined in the multimessenger picture, are the key to identifying the origins of CRs
and estimating transport properties. While gamma-ray observations alone raise the question of whether their origin is hadronic or leptonic, the observation of high-energy neutrino emission directly points to the presence of CR hadrons. 
To identify the emission signatures from acceleration and transport effects a proper modeling of those interactions in a transport framework is needed. Significant work has been done to tune the production cross sections to accelerator data and different models exist that put the exact evolution of the Monte-Carlo generated showers into a statistical approach of a probabilistic description of the production of the final states of the showers relevant for astrophysical observations. 

In this work, we present the implementation of different hadronic interaction (HI) models into the publicly available transport code CRPropa. 
We apply different descriptions of the HI, trained on observational data in different energy regimes to a nearby, giant molecular cloud \change{and the Galactic diffuse emission}. In this case, the resulting gamma-ray \change{and neutrinos fluxes} can differ by a factor $\sim 2$ dependent on the choice of the HI model.
}

\begin{document}
\maketitle
\flushbottom

\section{Introduction}
Hadronic interactions of Cosmic Rays (CRs) play an important role in the multimessenger interpretation of astrophysical environments that show signs of particle acceleration. Interactions of CRs with ambient photon fields or gas at the acceleration site are expected to give rise to non-thermal emission of high-energy photons and neutrinos. 
In radiation-dominated environments, electromagnetic and photohadronic interactions, such as $p\gamma \to pe^+e^-$ or $p\gamma \rightarrow p+\pi^0, n+\pi^- \rightarrow \gamma,\nu$ above the respective production thresholds, might dominate (see e.g.\ \cite{becker2008} for a review). However, in many astrophysical environments, hadron-hadron interactions play an important role, and a clear signature of such interactions is given by high-energy neutrinos resulting from the decay of pions and kaons in leading order.
%Those in turn decay into neutrinos and gamma rays.
Hadrons with strange and charm contents also exist, but are subdominant in environments that are transparent to pions and kaons, which is typically the case in astrophysical environments (see e.g.\ \citep{charmed_galaxies2023}). 

Gamma-ray observations by Fermi have by now revealed thousands of sources above GeV energies. In the TeV range, Imaging Air Cherenkov Telescopes (IACT) and air shower detectors could confirm more than 300 sources so far. \change{A special effort has been made to compile cataloges of known galactic and extragalactic sources in the GeV \cite{4FGL} and TeV energies \cite{TeVCat}.} Leptonic processes are expected to contribute significantly to these signatures, i.e.\ via inverse Compton scattering or bremsstrahlung by relativistic electrons. Hadronic emission via $\pi^0$ decay is also expected from a large number of source classes like jets and cores of active galaxies, starforming galaxies, and galaxy clusters for extragalactic objects, see e.g.\ \citep{becker2008} for a review. Due to the overlapping signatures from electrons and hadrons in gamma-ray observations in the GeV-TeV range, the signatures are ambiguous. Evidence for neutrino emission has been reported for the Seyfert galaxies NGC1068 \citep{icecube_ngc1068} and NGC4151 \citep{IceCube2024}, as well as for the blazars TXS0506+056 \citep{2018Sci...361..147I} and PKS1424+024 \citep{2018Sci...361.1378I}. 
For Galactic objects, evidence for the pion bump expected from $\pi^0 \to \gamma \gamma$ decay photons has been found for the supernova remnants W51, IC443, and W28 \citep{Fermi_W51,Fermi_IC443, Fermi_W28}. 
Other candidates for hadronic emission are microquasars, X-ray binaries, Pulsar Wind Nebulae, and more generally the Galactic Center \citep{Abramowski2016,hess_GC_18,delaTorre2023, Scherer_GC_2022, Scherer_GC_2023, Doerner2024}. 
A review of Galactic sources of cosmic rays and their interpretation in the multimessenger picture is found e.g.\ in \citep{beckertjus_merten2020}. Interactions of Galactic cosmic rays with the diffuse gas in the Milky Way also produce a significant amount of gamma rays \citep{Hunter1997,Abdo2008,2022SNAS....4...15R,Porter2017, HAWC_diffuse, Hermes}
and have recently been observed at extreme gamma-ray energies by LHASSO \citep{LHAASO_plane}, and in neutrinos \citep{IceCubePlane}. Filamentary clouds in our local neighborhood have been investigated in \cite{Youssef2024} to study the local diffusion coefficient of cosmic rays using gamma-ray observations. A difficulty in this approach lies indeed in the uncertainties of the hadronic cross sections. In this paper, we therefore use the setup of such a cloud as 
a test case in order to investigate the impact of the cross-section model used in the modeling.

The implementation of hadronic interactions in astrophysical modeling relies on knowing key features such as the cross sections, branching ratios, and produced particle multiplicities. Much information is provided from a combination of theoretical and experimental particle physics, today led by measurements at CERN, see e.g.\ \citep{2022Ap&SS.367...27A} for a review. 
But even the Earth's atmosphere has long been a laboratory in which the decay products of those showers can be detected to understand the development of the showers, the particle production at the interaction point as well as cross sections, see e.g.\ \cite{Fedynitch2012, PhysRevLett.109.062002}.
Accelerator and air shower data provide complementary information about the phase space of produced particles: Accelerator data provide precision information about particle production at large transverse momenta up to projectile energies of $\sim 10^{17}$~eV. In comparison, extensive air shower measurements are sensitive to the most forward kinematic region up to particle energies of $\sim 10^{20}$~eV. Combining both is therefore extremely useful to improve the description of particle shower development for the full phase space of momentum and direction (see e.g.\ \cite{2022Ap&SS.367...27A} for a review).

With the ongoing effort of combining cross sections, it is still a major challenge to consistently describe hadronic cross sections over the large energy range which is necessary when trying to model multimessenger data. This is usually done from GeV to PeV energies and beyond. For example, when studying hadronic interaction models to explain the gamma-ray emission from filamentary clouds, it was pointed out in \citep{Youssef2024} that large uncertainties in the interpretation of the gamma-ray observations arise from uncertainties in the hadronic cross sections. In particular, interaction codes are typically tuned either to cosmic-ray energies below or above $\sim 100$~GeV. The reason is that high-energy interaction codes like 
PYTHIA \cite{Pythia8.3}, SYBILL \cite{Riehn:2019jet}, QGSJET \cite{QGSJET-III}, and EPOS \cite{EPOS-LHC}
are tuned to LHC data at TeV energies. Often, they are combined with low-energy codes like FLUKA \cite{FLUKA} and UrQMD \cite{URQMD} that are optimized to describe fixed target experiments and previous accelerator data at lower energies.

To model gamma-ray and neutrino emission, the codes can generally be used to derive the probability distribution functions (PDFs) of the pions and kaons, decaying into electrons, gamma rays, and neutrinos. However, implementing the full interaction codes in the transport equations of astrophysical simulations is typically not feasible, as the codes are usually not based on a Monte Carlo approach and are difficult to combine with Monte Carlo generators. Contrary to other codes, CRPropa 3.2 \citep{Batista2022} is a Monte Carlo-based framework that -- besides simulating ballistic propagation of particles -- allows solving the transport equation via the approach of stochastic differential equations \citep{Merten:2017mgk}. The direct coupling of CRPropa to the interaction codes would require an extensive amount of computational time due to the probabilistic nature of the process. To speed things up, the event generators are used to derive PDFs for a discrete set in momentum space. The directional information of the scattering is usually averaged and isotropy of the astrophysical system is assumed. The resulting products from the hadronic interactions of cosmic rays, in particular, the decay products neutrinos and gamma rays can be obtained with good accuracy with such a simplified procedure, in zeroth order using the integral cross section $\sigma_{pp}(T_p)$ with $T_p$ as the kinetic energy of the incoming proton,  see e.g.\ \citep{2000A&A...362..937A}.
However, this approach makes it impossible to capture the impacts of the PDF resulting from the folding of the differential cross section in momentum and also scattering direction. The latter is typically integrated out as (a) at relativistic energies, forward scattering is a reasonable approach and (b) for many astrophysical problems, the assumption of isotropy is reasonable. The single-differential description of the cross section has been estimated in e.g.\ \citep{Kelner2006,Kafexhiu2014,AAfrag,ODDK22,ODDK23}. A different approach to include hadronic interactions by directly calling the hadronic codes within the framework of CRPropa \citep{Morejon2023} is also available, although with performance limitations depending on the scenario. Such an implementation is complementary to the one presented here, since it can be used for testing new generators and for producing updated interaction tables to replace the models used here.

In this paper, different existing simplified descriptions of the production of gamma rays and neutrinos are used to quantify the systematic error that is introduced from the broadband modeling of the cross sections. In Section \ref{crosssections:sec}, the different available descriptions are reviewed. In Section \ref{sec:crpropa}, their implementation in the propagation environment of CRPropa is described and Section \ref{tests:sec} presents the validation of the code. The different interaction models are applied in Section \ref{sec:application} to the test \change{the gamma ray emission from a} standard molecular cloud and the uncertainties in the modeling relative to the different approaches are quantified. \change{To demonstrate the uncertainties for the resulting neutrino flux the different cross section models are applied to the emission from the Galactic plane in Section \ref{sec:application_neutrino}.} In Section \ref{conclusions:sec} we summarize the results and present an outlook. This work is the starting point to build cross sections that selfconsistently range from the lowest relevant energy range of cosmic-ray interactions ($\sim 0.1-1$~GeV) up to $>$PeV energies in order to minimize systematic effects from the cross section modeling in future works.

\section{Descriptions of hadronic cross sections \label{crosssections:sec}}

The computation of hadronic interaction quantities, such as the cross section and distributions of secondaries, is a complex task. On one hand the stochastic nature of the process requires a probabilistic approach or Monte Carlo methods, on the other hand, the existing theoretical frameworks are limited to few parton scatterings and become very complex when collective effects like multiple scatterings are important due to non-linear terms. This situation implies that codes are required to achieve an adequate level of description of experimental data by combining the theories with effective descriptions whose on parameters are adjusted by comparison to measurements and extrapolate to energy and kinetic ranges not available. For example, QGSJET~\cite{QGSJET-III} is based on Reggeon Field Theory which treats soft and hard processes in the same way, thus it does not require an effective parameter regulating their relative contributions \citep{2013EPJWC..5202001O}. However, it's less suitable for central nucleus-nucleus collisions \citep{2013EPJWC..5202001O}. Sibyll~\citep{Riehn:2019jet} is based on the dual parton model and assumes partons in projectile and target become connected by strings whose fragmentation leads to the production of secondaries, which is modeled via the Lund string fragmentation model. In addition, it implements a minijet model to account for the prevalence of hard scatterings for high energies such leading to increase in multiplicity and large transversal moments of the secondaries, etc. These ingredients, however, also fail to describe interactions where a large number of partons are involved, and the model is limited to interactions of nuclei of low mass since it employs Glauber model for hadron-nucleus collisions.
Other hadronic interaction codes available like PYTHIA \cite{Pythia8.3} and EPOS \cite{EPOS-LHC} have different approaches, being tuned to describe data in accelerators and extensive air showers, however, they are not employed for the models described here.

The codes currently available are able to provide PDFs of the particle population at the interaction vertex and cross sections for individual processes. To receive these, extensive Monte Carlo simulations need to be run to cover the full phase space of the processes, which requires a significant amount of computation time. For cosmic-ray applications it is, therefore, more convenient to employ precomputed cross sections and PDF tables produced by pre-runs of the models mentioned above. The parametrizations for distributions and inelastic cross sections employed in these phenomenological models are obtained by fitting the output of the hadronic models.  In cosmic-ray propagation models like GALPROP \cite{Porter2017}, DRAGON \cite{Evoli2016}, PICARD \cite{Kissmann2014}, \change{USINE \cite{Maurin20}} or CRPropa \cite{Batista2022}, it is necessary to make use of such effective description in order to save computational time. Those effective models on the marked, however, have often been tuned on a certain energy range and never cover all energies from GeV to PeV energies necessary to make multimessenger studies. This is why we are presenting the systematic study of the uncertainties that are connected to the use of the different models.

\subsection{Total inelastic cross section} \label{ssec:inel_cross}
The total inelastic cross section for proton proton interactions has been measured across a large range of energies. A review including recent data is compiled by the Particle Data Group (PDG) \cite{data_pdg}, providing the total and elastic cross section in machine readable format\footnote{\url{https://pdg.lbl.gov/2022/hadronic-xsections/}}. Based on this, the inelastic cross section can be obtained as the difference between the total and the elastic one.  The theoretically motivated fit function for the inelastic cross section at high energies is quadratic in the logarithm of the proton energy. The authors introduce a cut-off at low energies to the model in order to reflect the kinematic threshold of the pion production. 

\begin{figure}
    \centering
    \includegraphics[width=\textwidth]{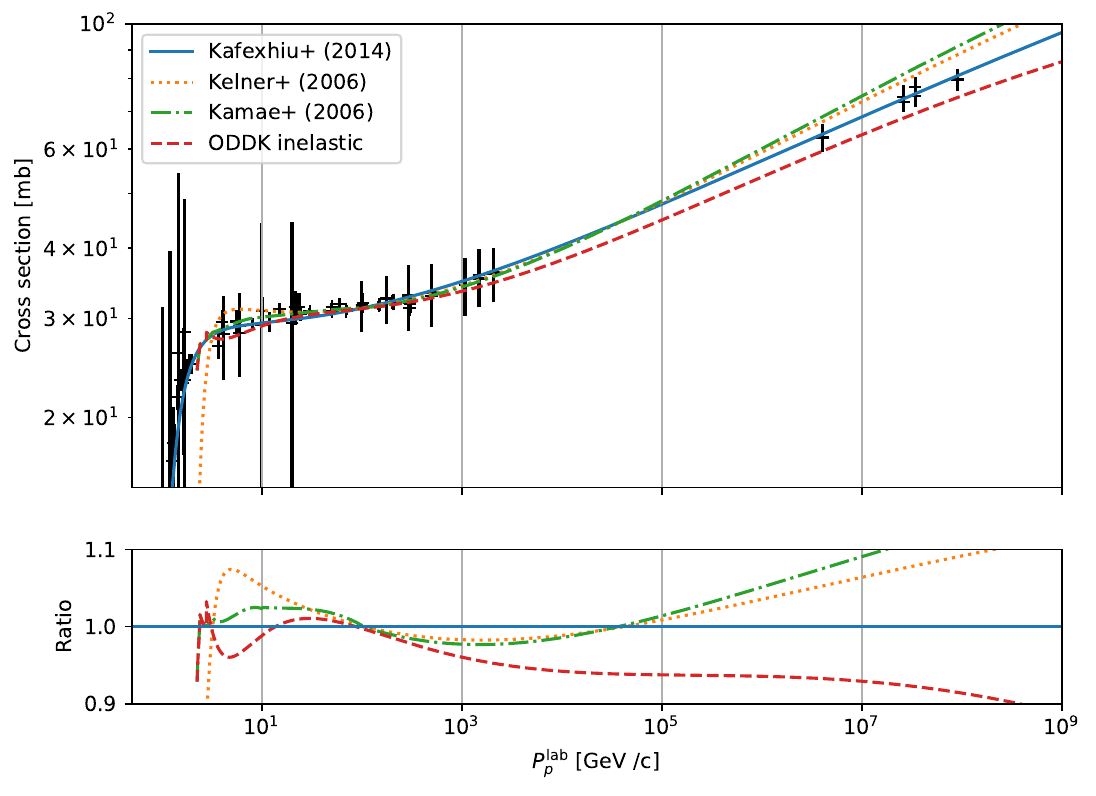}
    \caption{Comparison of the inelastic cross section parametrization \cite{Kafexhiu2014, Kelner2006, Kamae2006, ODDK22} with the data from the Particle Data Group \cite{data_pdg}.}
    \label{fig:inel_cs}
\end{figure}

In this work, we make use of the fit by Kafexhiu \etal \cite{Kafexhiu2014}
given by 
\begin{equation}
    \sigma_\mathrm{inel}(T_p) = \left[ 30.7 -0.96 \, \log(r) + 0.18 \log^2(r)\right] \cdot \left[1 - r^{-1.9}\right]^3 \ \mathrm{mb}\label{eq:inelastic} \quad , 
\end{equation}
using $r = T_p / T_p^\mathrm{th}$ as the ratio between the kinetic energy $T_p$ and the threshold energy $T_p^\mathrm{th} = 2 m_\pi + m_\pi^2 / 2m_p \approx 0.2797 \ \mathrm{GeV}$.
In Fig.~\ref{fig:inel_cs} the fit of the inelastic cross section is compared to measurements calculated from the total and elastic pp cross section data by the PDG \cite{data_pdg} and \change{different fits from \cite{Kelner2006, Kamae2006, ODDK22}.}
The fit by Kafexhiu \etal \cite{Kafexhiu2014} is in better agreement with the measurements at the highest energies than the fits from \change{other works. It should be noted that the work of Orusa \etal \cite{ODDK22} fits the total and the elastic cross section and calculates the inelastic as their difference. This fit takes into account more recent data compared to the work of Kafexhiu \etal and achieves a good agreement. The amount of data for the inelastic measurements shown in Fig.\ \ref{fig:inel_cs} is much less. In the following, we apply the parametrization of Kafexhiu \etal \cite{Kafexhiu2014}, but we note that this choice introduces additional uncertainties in the resulting fluxes.}

\subsection{Differential inclusive cross sections} \label{ssec:diff_cross}
The measurements by collider experiments and the full interaction codes provide the fully differential production cross section, typically stated in the Lorentz invariant form:
\begin{equation}
    \sigma_\mathrm{inv} = E_s \, \frac{\diff^3 \sigma}{\diff p_s^3} = E_s \frac{\diff \sigma}{p_s^2 \, \diff p_s \, \diff \Omega} \, .
\end{equation}
Here, $E_s$ is the total energy of the secondary particle and \change{$p_s$} is its momentum.
In astrophysical applications, where the interactions can be described as a fixed target scattering, it is useful to fit the cross section $\diff \sigma / \diff p$ integrated over the solid angle $\Omega$. Depending on the model, the fit is performed in dependence on the momentum $p_s$, the kinetic energy $T_s$, or the total energy $\epsilon_s$ for the secondary species $s$. 

\begin{figure}
    \centering
    \includegraphics[width=\linewidth]{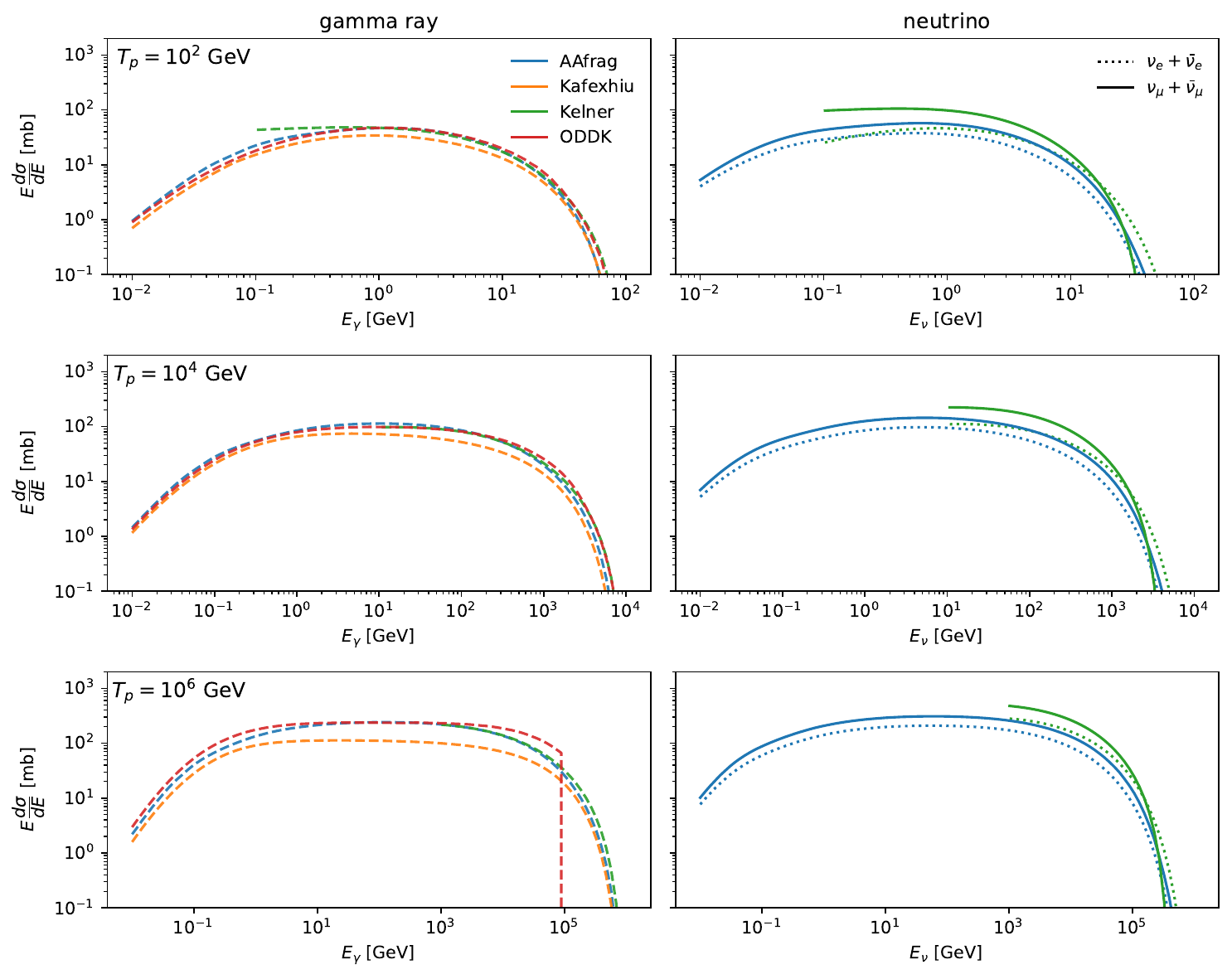}
    \caption{\change{Differential cross section  \cite{AAfrag,Kafexhiu2014,Kelner2006,ODDK22, ODDK23} for gamma rays (left column) and neutrinos (right column) at different kinetic energies of the primary (row).}}
    \label{fig:diff_crosssection}
\end{figure}

In the following, we discuss different proposed parametrizations, which have been included in the CRPropa framework in the context of this paper (see Section \ref{sec:crpropa} for the implementation). \change{A direct comparison of the differential cross section models at different primary energies is given in Fig.\ \ref{fig:diff_crosssection}.} 

\paragraph{Kelner \etal 2006}
The model by Kelner \etal\cite{Kelner2006} describes the spectra $F_{s}$ of secondary species, $s$, including pions, gamma rays, electrons, and neutrinos. It is based on an analytical fit of the PDFs resulting from simulations with SIBYLL \cite{SIBYLL}. The model is based on sampled interactions with kinetic energy $T_p$ in the range of $0.1 - 10^5$ TeV and is fitted for the ratio between the secondary and primary energy $x = E_\mathrm{sec} / E_\mathrm{prim} \gtrsim 10^{-3}$. 
In the production of pions, it is assumed that $\pi^{-}$ and $\pi^{+}$ are produced in same numbers because of isospin conservation, resulting in the same number of muons and antimuons as well as a first generation of muon neutrinos $\nu_{\mu}^{1}$. The muons  decay to $\mu^{-}\rightarrow e^{-}\,\overline{\nu}_e\,\nu_{\mu}$ and $\mu^{+}\rightarrow e^{+}\,\nu_e\,\overline{\nu}_{\mu}$. The PDF of $e^{\pm}$ and $\nu_{e}$ are the same when neglecting the mass of the electron and only the muon (anti) neutrinos from the muon decay as well as the first generation muon neutrinos each receive a different spectrum.   

For the lower energy extension ($T_p \leq 100$ GeV) the authors introduce a so called \textit{$\delta$-approximation} (see \cite{Schlickeiser02} for a more detailed discussion), where the spectrum is parameterized as 
\begin{equation}
    F_s = \tilde{n} \, \delta\left(\epsilon_s - \frac{\kappa}{\tilde{n}} T_p\right) \quad, 
\end{equation}
using the average number of secondaries $\tilde{n}$ per interaction and the fraction $\kappa$ transferred into the corresponding secondary channel. The parameters for $\tilde{n}$ and $\kappa$ are based on the assumption of the continuity of the spectrum at 100 GeV. Therefore, this approach can not be used in the description of a single particle interaction, as the spectral slope is not known beforehand.

Knowing the inelastic cross section $\sigma_\mathrm{inel}$ as given in section \ref{ssec:inel_cross}, the differential cross section for the species $(s)$ can be derived as 
\begin{equation}
    \frac{\diff \sigma^{(s)}}{\diff \epsilon}(T_p, \epsilon) = \frac{\sigma_\mathrm{inel}(T_p)}{T_p} \, F_{(s)}(T_p, \epsilon / T_p) \quad .
\end{equation}

\paragraph{Kafexhiu \etal 2014}
Kafexhiu \etal\cite{Kafexhiu2014} expand the parametrization for secondary gamma rays arising from p-p interactions. Neutrinos, electrons and positrons are not included. In their paper, the authors focus on the lower energies, including different Monte Carlo generators such as GEANT 4 \cite{GEANT4}, PYTHIA 8.1 \cite{PYTHIA81}, and SIBYLL 2.1 \cite{SIBYLL21}, and published data for interactions with kinetic energies below 2 GeV. At the lowest energies ($T_p < 2$ GeV) the production of secondary gamma rays includes baryon resonances. The parametrizations are fitted for a primary energy range $T_p^\mathrm{th} \leq T_p \leq 1 \, \mathrm{PeV}$, starting at the kinematic threshold of the p-p interaction $T_p^\mathrm{th} \approx 0.2797$ GeV. The results from each event generator and the observed data are fitted independently. 

In this work, we use the fit values based on data at the lowest energies ($T_p \leq 1 \, \mathrm{GeV}$), GEANT in the intermediate energies ($T_p \leq 50 \, \mathrm{GeV}$) and PYTHIA at the highest energies ($T_p > 50 \, \mathrm{GeV}$). For the implementation as a CRPropa module (see section \ref{ssec:pre-data}) the user can easily change the values.

\paragraph{AAfrag}
The AAfrag model \cite{AAfrag} is based on simulations done with QGSJET-II-04m \cite{QGSJET-II}. In that work, the authors do not only focus on light secondaries, such as electrons ($e$), positrons ($e^+$), neutrinos ($\nu_e, \nu_\mu$), and gamma rays from the pion channels, but also include secondary hadronic channels. Those channels include secondary protons ($p$) and neutrons ($n$) as well as their antiparticles. 

Additionally, the authors investigate the difference in the interaction between different projectiles and targets. The projectiles cover examples from the typical CR mass groups, including protons, helium, carbon, aluminum, and iron. 
For all projectile species, the interactions are sampled using a proton target. Additionally, for proton and helium primaries, also a helium target was considered.

The lower boundary of the tested energy range varies with the considered primary particle, starting at $T_p = 5 \, \mathrm{GeV}$ for protons and going up to $T_p = 100 \, \mathrm{GeV}$ for iron. The authors refer to a maximum energy of $T_p = 10^{20} \, \mathrm{eV}$. It should be noted that accelerator data do not reach those energies and reliable results can only be expected up to $\sim 10^{17}$~eV.

\paragraph{ODDK}
The model ODDK by Orusa \etal\cite{ODDK22, ODDK23} focuses on the fluxes of electrons, positrons ($e^\pm$), and gamma rays ($\gamma$) from Galactic CRs. The authors focus on all production channels contributing to the secondary species at least 0.5\% of the total yield. For the leptonic channel this includes the production of pions ($\pi^\pm \rightarrow e^\pm + X$ and $\pi^0 \rightarrow e^+ + e^- + \gamma$) as well as kaons ($K^+, K^-, K_s^0, K_l^0$) and lambda baryons ($\Lambda, \bar{\Lambda}$). In the gamma ray channel, the contribution from $\eta$ mesons ($\eta \rightarrow 2\gamma$, $\eta \rightarrow 3\pi^0 \rightarrow 6\gamma$ and $\eta \rightarrow \pi^+\pi^-\pi^0 \rightarrow 2\gamma$) is covered as well.

The authors provide online tables for the electron and positron\footnote{\url{https://github.com/lucaorusa/positron_electron_cross_section}} and gamma-ray\footnote{\url{https://github.com/lucaorusa/gamma_cross_section}} production cross sections. Tables for the neutrino production are not provided. The electron and positron tables cover the primary energy range $0.1 \, \mathrm{GeV} \leq T_p \leq 10^{6} \, \mathrm{GeV}$ and the secondary energy $10^{-2} \, \mathrm{GeV} \leq \epsilon_{e^\pm} \leq 10^{4} \, \mathrm{GeV}$. The gamma ray table is provided for $0.1\, \mathrm{GeV} \leq T_p \leq 10^7 \, \mathrm{GeV}$ and $10^{-2}\, \mathrm{GeV} \leq \epsilon_\gamma \leq 10^{5} \, \mathrm{GeV}$. In this work, we rely only on those parts of the tables that cover secondary energies up to the maximal kinetic energy of the primary $\epsilon = T_p$. Therefore, the maximum energy of the process is limited to $T_p^{(\gamma)} = 10^5 \, \mathrm{GeV}$ and $T_p^{(e^\pm)} = 10^{4} \, \mathrm{GeV}$. 

To cover the impact of heavier projectiles and targets, the authors derive a scaling function 
\begin{equation}
    \sigma_\mathrm{inv}^{A_1 A_2} = f^{A_1 A_2} \sigma_\mathrm{inv}^{pp}
\end{equation}
for the Lorentz invariant cross section using three fit parameters. Based on this scaling the provided tables cover proton and helium as a target and $^1\rm H$, $^2\rm  H$, $^3\rm  He$, $^4\rm  He$, $^{12}\rm  C$, $^{13}\rm  C$, $^{14}\rm  N$, $^{15}\rm N$ and $^{16}\rm O$ as projectile. 

\section{CRPropa implementation} \label{sec:crpropa}
CRPropa 3.2 \citep{Batista2016, Batista2022} is a Monte Carlo tool for the propagation of high- and ultra-high energy cosmic rays. Its modular structure allows for easy customization of the simulation and extension for new physical processes. In this section, we describe how the hadronic interactions are implemented as an additional module. The work is based on the plug-in template provided by CRPropa and is publicly available\footnote{\url{https://gitlab.ruhr-uni-bochum.de/doernjkj/hadronic-interaction-in-crpropa}}.

\subsection{Pre-calculated data} \label{ssec:pre-data}
To run the simulation, a set of tabulated data is needed. 
The data contain the cumulative distribution function $cdf$ of a given secondary species $s$ for any given kinetic energy of the primary $T_p$ and secondary $\epsilon$. The unnormalized $cdf$ is calculated by the integral of the differential cross section
\begin{equation}
    cdf^{(s)}(T_p, \epsilon) = \int\limits_{0}^{\epsilon} \mathrm{d}\epsilon^\prime \, \frac{\mathrm{d}\sigma^{(s)}}{\mathrm{d} \epsilon^\prime}(T_p, \epsilon^\prime) \, .
\end{equation}
In the case of models that do not provide all possible secondaries, an energy loss correction factor $f_\mathrm{loss}^{(s)}$ is calculated. This factor corrects the missing energy loss due to non-included channels. 
It scales the total energy loss of a secondary channel $s$ by the factor $f_\mathrm{loss}^{(s)}$.

For all models described in Section \ref{ssec:diff_cross}, the data are included in the plug-in. Also, a script is provided to calculate new tables for any given cross section and species.

\subsection{Working principle}
In each step of the propagation, CRPropa calls the \texttt{process} function of all included modules. For the hadronic interaction module, the process function evaluates whether an interaction should happen within the step. To decide on the interaction, the module takes the particle type and the interaction probability, calculated as 
\begin{equation}
    p = n_\mathrm{nucl} \cdot \sigma_\mathrm{inel} \cdot \Delta x \, , \label{eq:prob}
\end{equation}
into account. 
Here, $n_\mathrm{nucl} = n_{H_I}(\Vec{r}) + 2 \, n_{H_2}(\Vec{r})$ is the total nucleon number density of the atomic and molecular hydrogen target plasma at the position of the \texttt{Candidate}, $\sigma_\mathrm{inel}$ is the inelastic cross section following Eq.\ \eqref{eq:inelastic}, and $\Delta x$ is the current propagation step. We note this is only a first-order approximation of the exponential for the optical depth and only valid for small step sizes, but this assumption is made for all interactions within CRPropa, which also assumes a constant density over the full step. By comparing the calculated interaction probability with a random number, the decision if an interaction happens is made. 

In case of no interaction, the next propagation step is limited to a fraction $f$ of the mean free path. The default is $f = 0.1$, but can be changed by the user. 

In case of an interaction, the function \texttt{performInteraction} is called. This function performs a loop over all included secondaries. 
The first step is to calculate to total number of allowed secondaries 
\begin{equation}
    N_\mathrm{sec} = \frac{1}{\sigma_\mathrm{inel}} \int\limits_{0}^{T_p} \mathrm{d}\epsilon \, \frac{\mathrm{d}\sigma}{\mathrm{d} \epsilon} \, .
\end{equation}
Afterwards, the allowed number of secondaries is sampled from the tabulated $cdf$ and added to the simulation. 
In the end, the total energy of the secondaries, corrected by the energy loss factor $f_\mathrm{loss}^{(s)}$, is subtracted from the primary. The resulting energy of the primary is determined by
\begin{equation}
    E_\mathrm{new} = E_0 - \sum\limits_{s} f_\mathrm{loss}^{(s)}\cdot \sum_{i = 0}^{N_\mathrm{sec}^{(s)}} \epsilon_i  \, .
\end{equation}
The cross section in the AAfrag model \cite{AAfrag} includes secondary protons and other nucleons. In this case, the resulting primary is already covered in the distribution of secondary protons. Therefore, the user can decide to deactivate the \texttt{Candidate} after the interaction. 

\begin{figure*}
    \centering
    \includegraphics[width=.9\textwidth]{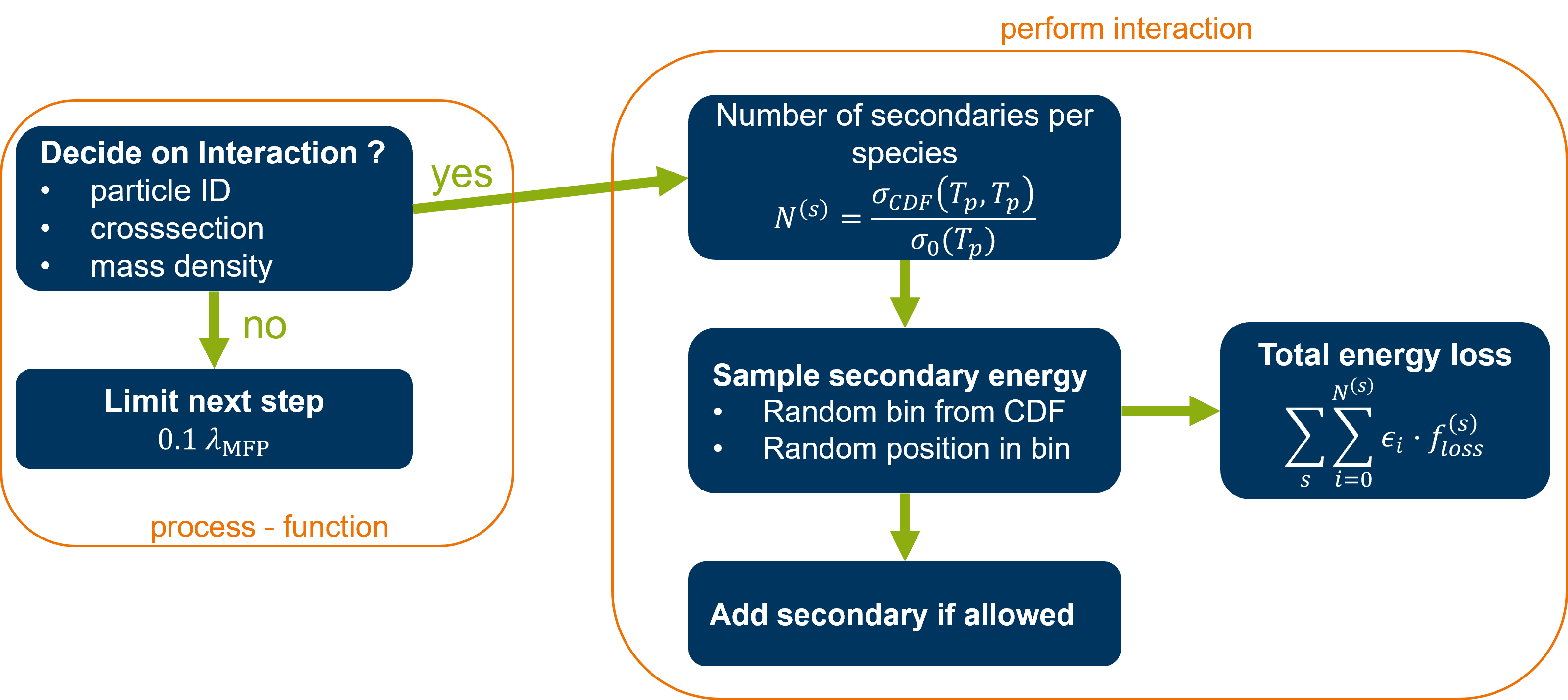}
    \caption{Workflow principle of the HI module. The start of the workflow is in the \texttt{process} function on the upper left.}
    \label{fig:workflow}
\end{figure*}

In Fig.\ \ref{fig:workflow}, the workflow of the HI module is shown. It shows the different sub-functions \texttt{process} and \texttt{performInteraction}.

\section{Validations \label{tests:sec}}
In this section, several first tests for the implementation of HI is given. The tests cover the mean free path (Sec.~\ref{ssec:test_mfp}), the secondary yields (Sec.~\ref{ssec:test_yields}), and the energy loss per time (Sec.~\ref{ssec:test_loss}). The test of the mean free path and the average energy loss are also included in the unit tests, which can be run in the installation process. 

\subsection{Mean free path} \label{ssec:test_mfp}
\begin{figure}
    \centering
    \includegraphics[width=\columnwidth]{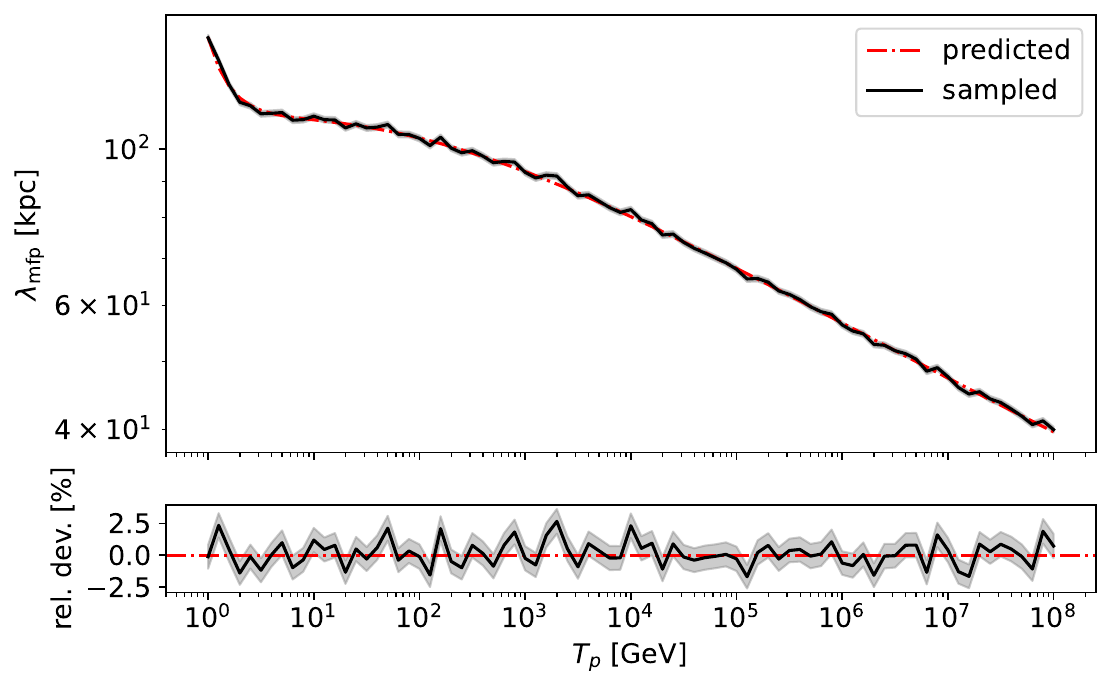}
    \caption{Comparison between the analytically expected mean free path and the one sampled from a simulation. The gray band indicates the standard deviation of the distance to the first interaction point.}
    \label{fig:MFP}
\end{figure}

As a first step, the mean free path of CRs in the simulation is tested. This test is designed to cross-check the evaluation of Eq.~\eqref{eq:prob}. For 81 primary energies in the range $1 \leq E / \mathrm{GeV} \leq 10^8$, we propagate $10^4$ primary protons in steps of $100$~pc in a medium with $n_{H_I} = 100 \, \mathrm{cm^{-3}}$ until the first interaction happens. In Fig.\ \ref{fig:MFP} the average distance to the first interaction in the simulations is shown and compared with the analytical expectation $\lambda_\mathrm{mfp} = (n\ \sigma_\mathrm{inel})^{-1}$. Only minor deviations due to the Monte-Carlo nature of the simulation can be seen. The error, indicated by the gray band, is calculated from the standard deviation of the mean $\sigma / \sqrt{n}$.  

\subsection{Yields} \label{ssec:test_yields}
In the second step, the multiplicity of the produced secondaries is tested. Here, we only refer to the case of the cross-section model AAfrag \cite{AAfrag}. The same test is performed for all models described in Section \ref{ssec:diff_cross} and are reported in Appendix \ref{app:yields}. 

\begin{figure*}
    \centering
    \includegraphics[width=\textwidth]{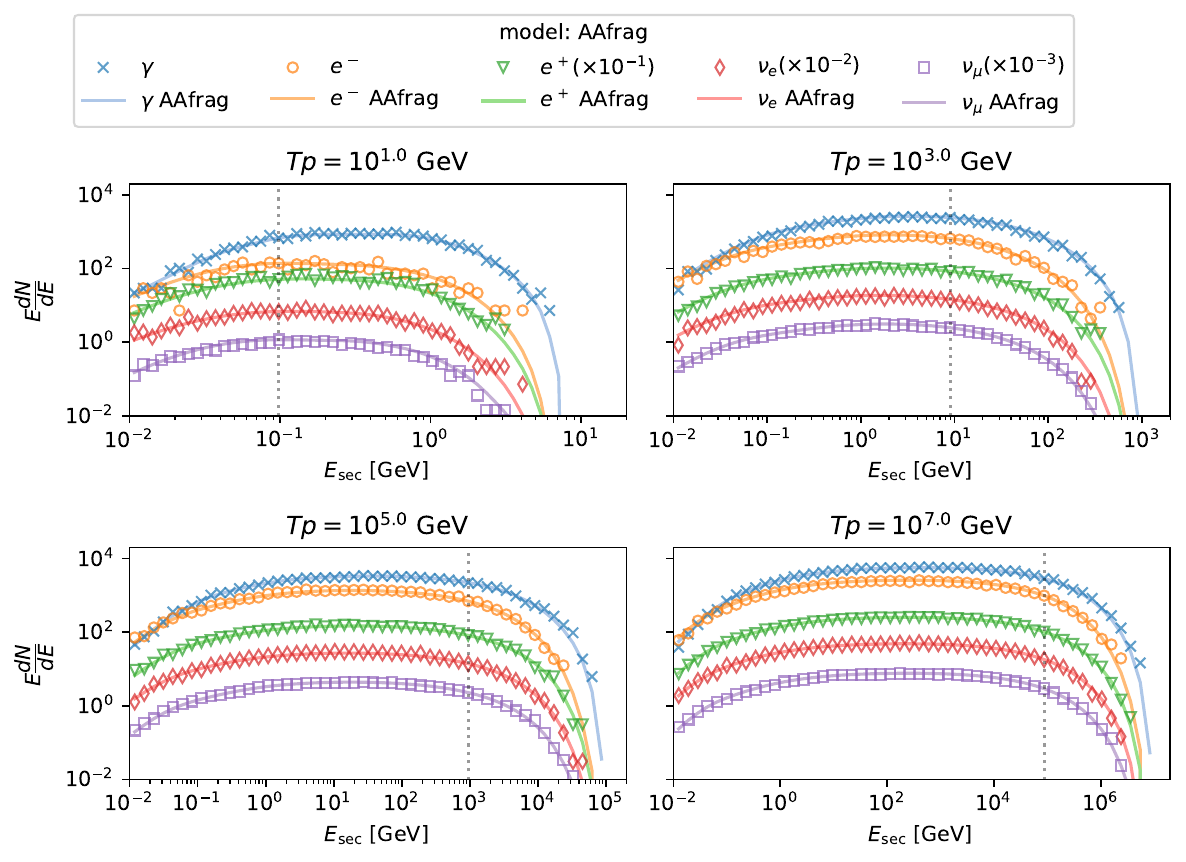}
    \caption{\change{Spectral yields of produced secondaries for the cross section model from \cite{AAfrag}.
    The symbols represent the sampled distribution from the module for different secondaries. The lines denote the predicted yield from the cross section model. For better visibility the positrons and neutrinos have been scaled down.}
    }
    \label{fig:yield_AAfrag}
\end{figure*}

To test the yields of the module, we perform $500$ interactions for a primary energy $T_p \in \left\{ 10^1, 10^3, 10^5, 10^7\right\} \, \mathrm{GeV}$. The resulting spectra of produced $\gamma$ rays, electrons ($e^-$), positrons ($e^+$) and neutrinos ($\nu_e$, $\nu_\mu$) is shown in Fig.~\ref{fig:yield_AAfrag}. Also, the predicted shape from the differential cross section is shown. The prediction is normalized at $E_\mathrm{sec} = 10^{-2}\, T_p$ to the sampled spectra, as this test is designed to check the spectral shape. The normalization is tested with the average energy loss in Section \ref{ssec:test_loss}. 

In general, the simulation is in good agreement with the predictions. Only for $E_\mathrm{sec} \approx T_p$ small deviations can be seen, which can be explained by the limited statistics and the Monte-Carlo nature of the code. 

\subsection{Energy loss rate} \label{ssec:test_loss}
The energy loss of protons due to hadronic interactions can be estimated by 
\begin{equation}
    - \frac{\diff E}{\diff t} (T_p) = \int\limits_{E_{th}}^{T_p} \diff \epsilon \ v \, \epsilon\, n(\Vec{r}) \ \sum\limits_{s} \frac{\diff \sigma^{(s)}}{\diff \epsilon} (T_p, \epsilon) \ , 
\end{equation}
where $n(\Vec{r})$ is the nucleon density at position $\Vec{r}$, $v$ the particle velocity, $s$ the secondary species, and $\sigma$ the inclusive inelastic cross section. This energy loss has been analytically approximated by \cite{Krakau2015} based on the inelastic cross section for charged and neutral pions  ($s = \pi^0, \pi^+, \pi^-$) described in \cite{Kelner2006}. Typically, this approximation is implemented in Galactic propagation codes, such as DRAGON2 \cite{Evoli2016}. The authors derive 
\begin{equation}
    -\frac{\diff E}{\diff t} \approx 3.85\cdot 10^{-16} \, \frac{n(\Vec{r})}{\mathrm{cm^{-3}}} \, \left( \frac{E}{\mathrm{GeV}}\right)^{1.28} \, \left(\frac{E}{\mathrm{GeV}} + 200 \right)^{-0.2} \, \frac{\mathrm{GeV}}{\mathrm{s}} \ . \label{eq:energyloss}
\end{equation}

To test the energy loss, we propagate $10^4$ protons for one step with $\Delta x = 0.01 \lambda_\mathrm{mfp}$ for each energy in a constant density of $n = 100~\mathrm{cm^{-3}}$. By comparing the energy after the propagation step $E_1$ with the initial energy $E_0$ we can estimate the energy loss as 
\begin{equation}
    - \frac{\diff E}{\diff t} \approx \frac{E_0 - E_1}{\Delta x / c} \ .
\end{equation}
In the ODDK model, the total energy loss in leptons ($e^\pm$, $\nu$) is assumed to be $f_\mathrm{loss}^{(e^\pm)} = 4$ times the total energy of produced electrons and positrons. This assumption is necessary as the ODDK model does not provide any cross section for  neutrinos. The factor can be followed from the dominant $\pi^\pm$ decay where in total four leptons (one electron/positron and three neutrinos) are produced per interaction.

\begin{figure}
    \centering
    \includegraphics[width=\linewidth]{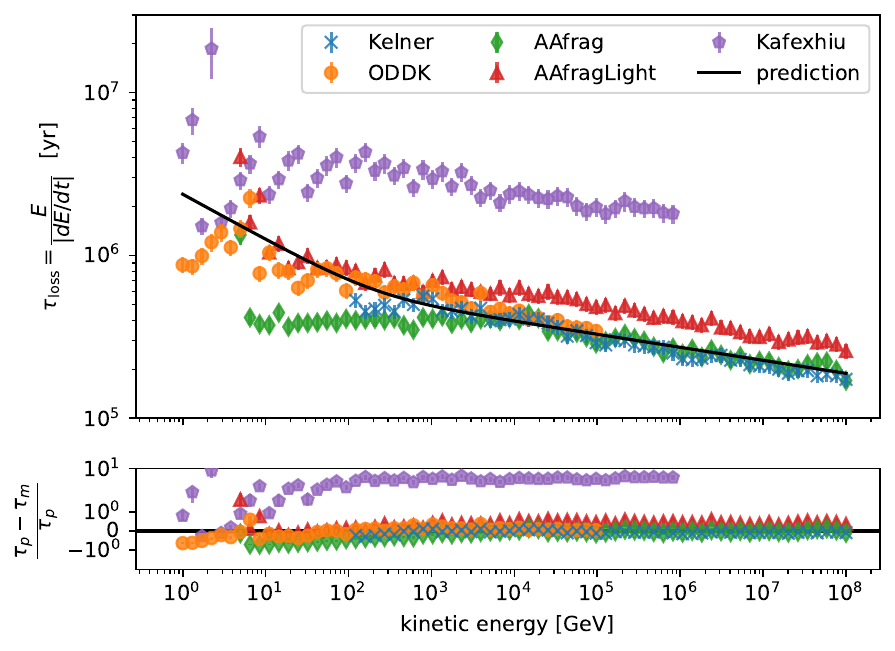}
    \caption{Estimated energy loss time for different cross section models in a constant density of $n = 100 \, \mathrm{cm^{-3}}$. The prediction is based on the approximation by \cite{Krakau2015} given in Eq.\ \eqref{eq:energyloss}. The \textit{AAfragLight} model only including light particles ($\gamma$, $e^\pm$, and $\nu$) which are included in the other models. The lower panel shows the relative deviation between the sampled energy loss timescale $\tau_m$ and the prediction $\tau_p$.}
    \label{fig:energyloss}
\end{figure}

In Fig.\ \ref{fig:energyloss} the loss timescale $\tau_\mathrm{loss} = E / | \mathrm{d}E / \mathrm{d}t|$ is shown based on the sampled energy loss. The data for the \textit{Kelner} model (blue crosses) show a good agreement to the analytical approximation as they are derived from the same model, but assuming different secondary species. 

The \textit{AAfrag} model (green diamond) agrees with the high energy part of the \textit{Kelner} model but has a significantly higher energy loss, resulting in lower loss timescales at lower energies ($E \leq 10^3 \, \mathrm{GeV}$). The \textit{AAfragLight} model (red triangles) contains only the light secondaries as gamma rays, electrons, positrons, and all flavors of neutrinos. The comparison between the full and the light model shows that a fraction of the total energy loss goes into hadronic secondaries. At the higher energies ($E > 10^3$ GeV), it is a constant factor. In the lower energy part, the total energy loss is more dominated by the hadronic contribution. Up to now, the \textit{AAfrag} model is the only description containing secondary hadrons. 

The \textit{ODDK} model (orange circles) predicts a similar energy loss as the \textit{Kelner} model. This agreement allows a validation of the energy loss factor $f^{(e^\pm)}_\mathrm{loss} = 4$ to account for the missing contribution of neutrinos. We note that this treatment is only valid on average of a high number of particles but not for individual interactions. 

The \textit{Kafexhiu} model (purple pentagons) shows a significantly larger energy loss timescale than the other models. This may be expected as this model only contains secondary gamma rays. For a real application, one would need to assume a cross section for the leptonic ($e^\pm$, $\nu$) emission, which could be taken from other models presented in Section \ref{ssec:diff_cross}. 
As the energy scaling of the average energy loss strongly differs from the other models, a simple scaling factor, as applied in the \textit{ODDK} model is not possible.

\section{Test application to the general setup of giant molecular clouds \label{sec:application}}
To illustrate the application of the CRPropa module we investigate the gamma-ray production in a simplified, generic giant molecular cloud (GMC) from the interaction of cosmic rays from the local interstellar spectrum. 
The gamma-ray emission from some local GMC close to the heliosphere show an excess than the expectation from the CR spectrum measured at earth, while other clouds are in agreement with the expectation \cite{Baghmanyan2020}. Depending on the model approach, other studies do not find differences \cite{Roy23}. Therefore, we test the systematic uncertainty coming from the cross section model in a generic GMC. 

The cloud is modeled as a sphere with radius $R_\mathrm{GMC} = 10 \ \mathrm{pc}$ and a spherical symmetric density profile
\begin{equation}
    n(r) = \frac{n_0}{1 + \frac{r}{0.5 \, \mathrm{pc}}} \ ,
\end{equation}
with an absolute normalization of $n_0 = 10^{3} \, \mathrm{cm}^{-3}$ in the center. 
% The resulting profile is shown in Fig.\ \ref{fig:density}.

% \begin{figure}
%     \centering
%     \includegraphics[width=.6\columnwidth]{figures/density_profile.pdf}
%     \caption{Density profile of the synthetic GMC used for the simulation.}
%     \label{fig:density}
% \end{figure}

\subsection{Simulation setup}
To simulate the interactions of the CRs from the Local Interstellar Spectrum in the cloud we propagate $N_\mathrm{sim} = 10^{8}$ particles starting on a sphere around the cloud, with isotropic ingoing direction. The propagation is carried out on straight paths with an adaptive step length between $l_\mathrm{min} = 10^{-6}\ \mathrm{pc}$ and $l_\mathrm{max} = 0.1 \ \mathrm{pc}$ to ensure a step size smaller than the changes in density. Assuming a propagation on straight lines neglects all effects of interactions with a potential magnetic field in the cloud, which is believed to occur in these clouds, but it allows us to focus on the impact from the cross-section. 
\change{For the CRPropa simulation, we assume a flat energy distribution in log-space and then re-weight the propagated particles to a more realistic injection spectrum. This allows statistical uncertainties to be minimised without excessive computational effort.}
The details of the used CRPropa modules and parameters are summarized in Table \ref{tab:CRPropaModules}.

\change{The first case of the reweighting assumes the local interstellar spectrum of CRs, which is a reasonable assumption for a GMC in the local neighborhood of the heliosphere. The second case assumes a power-law injection, which can be expected for a GMC close to an acceleration site like a supernova remnant. Both analyses use the same simulation output. From} the simulation, a histogram of the detected gamma rays is calculated. We use $80$ logarithmic energy bins between $10^{-1}$ GeV and $10^{7}$ GeV. 
For each candidate a weight 
\begin{equation}
    w_i = \frac{j_p(E_0)}{E_0^{-1} \, N_\mathrm{sim}}
\end{equation}
is applied, where $E_0$ is the source energy and $j_p$ is the \change{corresponding injection spectrum.} For the local interstellar spectrum we are using the parameterized from \cite{Roy23} with 
\begin{equation}
    j_p(E_p) = 2.7 \, E_p^{1.12} \, \beta^{-2} \left( \frac{E_p + 0.67}{1.67} \right)^{-3.93}  \left[ 10^{-3}/(\mathrm{GeV\,m^2\,s\,sr}) \right] \label{eq:LIS} \ , 
\end{equation}
\change{while the power-law injection follows:}
\begin{equation}
    j_p(E_p) \propto \left( \frac{E}{1 \, \mathrm{GeV} }\right)^{-2.2} \ .
\end{equation}

\renewcommand{\arraystretch}{1.4}
\begin{table*}[tp]
    \centering
    \begin{tabular}{|c|c|} \hline 
         module & Parameters  \\
         \hline
         \multicolumn{2}{|c|}{\textbf{propagation}} \\ \hline
         \texttt{SimplePropagation} & $l_\mathrm{min} = 10^{-6} \ \mathrm{pc}$, $l_\mathrm{max} = 0.1 \ \mathrm{pc}$ 
         \\ \hline
        \multicolumn{2}{|c|}{\textbf{interaction}} \\ \hline
        \texttt{HadronicInteraction} & cross-section model, density profile \\
        \texttt{NuclearDecay} $^a$& \texttt{havePhotons = True} \\ \hline
        \multicolumn{2}{|c|}{\textbf{observer \& output}} \\ \hline
        \texttt{TextOutput} & \texttt{Event3D} \\
        \texttt{ObserverDetectAll} & \\
        \texttt{ObserverNucleusVeto} & \\ \hline
        \multicolumn{2}{|c|}{\textbf{source}} \\ \hline
        \texttt{SourceParticleType} & $(A,Z) = (1,1)$ \\ 
        \texttt{SourcePowerLawSpectrum} & $E_\mathrm{min} = 1 \, \mathrm{GeV}$, $E_\mathrm{max} = 10^7 \, \mathrm{GeV}$, $\alpha = -1$ \\
        \texttt{SourceLambertDistribuionOnSphere} & $R_\mathrm{GMC}$, $\Vec{r}_0 = (0,0,0)$, \texttt{inwards = True} \\ \hline
        \multicolumn{2}{|c|}{\textbf{boundary}}
        \\ \hline
        \texttt{MinimumEnergy} & $E_\mathrm{br} = 0.1 \ \mathrm{GeV}$ \\
        \texttt{MaximumTrajectoryLength} & $D_\mathrm{max} = 2.4 \, R_\mathrm{GMC}$
        \\ \hline       
    \end{tabular} \\ \ \\ 
    \footnotesize{$^a$ The nuclear decay is included for the AAfrag cross-section model, which also provides secondary neutrons, which can decay further.}
    \caption{CRPropa Modules used for the simulation of the synthetic GMC.}
    \label{tab:CRPropaModules}
\end{table*}

\subsection{Results}
\begin{figure*}[tp]
    \centering
    \includegraphics[width=\textwidth]{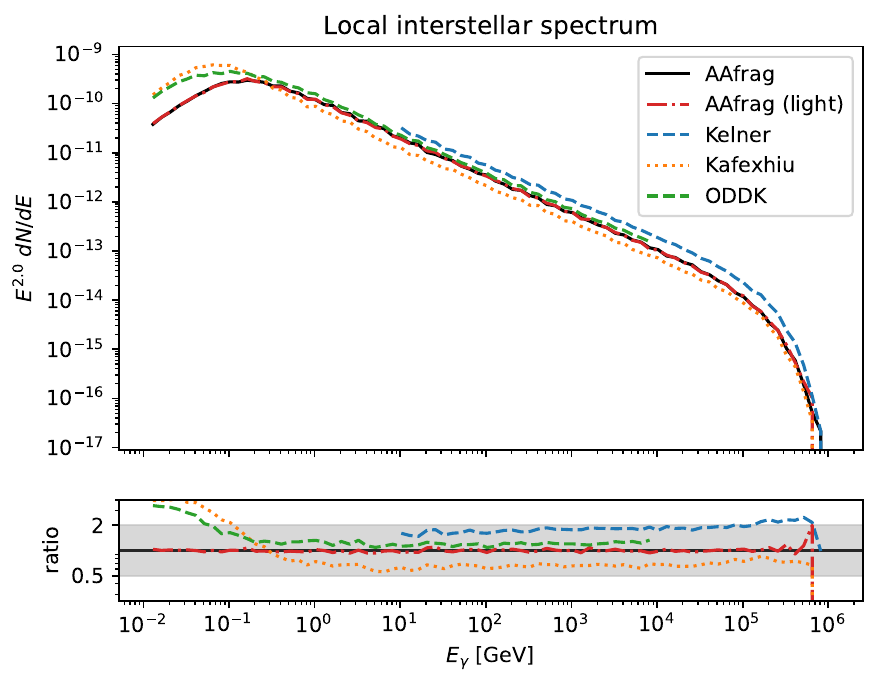}
    \caption{\textit{Upper:} Gamma-ray flux from the synthetic GMC based on different cross-section models. Here the injection spectrum from the local interstellar spectrum (eq.~\ref{eq:LIS}) is used.
    \textit{Lower:} ratio between the different flux predictions. Here, the AAfrag model is used as a baseline. The gray band indicates a difference of a factor 2.}
    \label{fig:gamma_flux}
\end{figure*}

\begin{figure*}[tp]
    \centering
    \includegraphics[width=\textwidth]{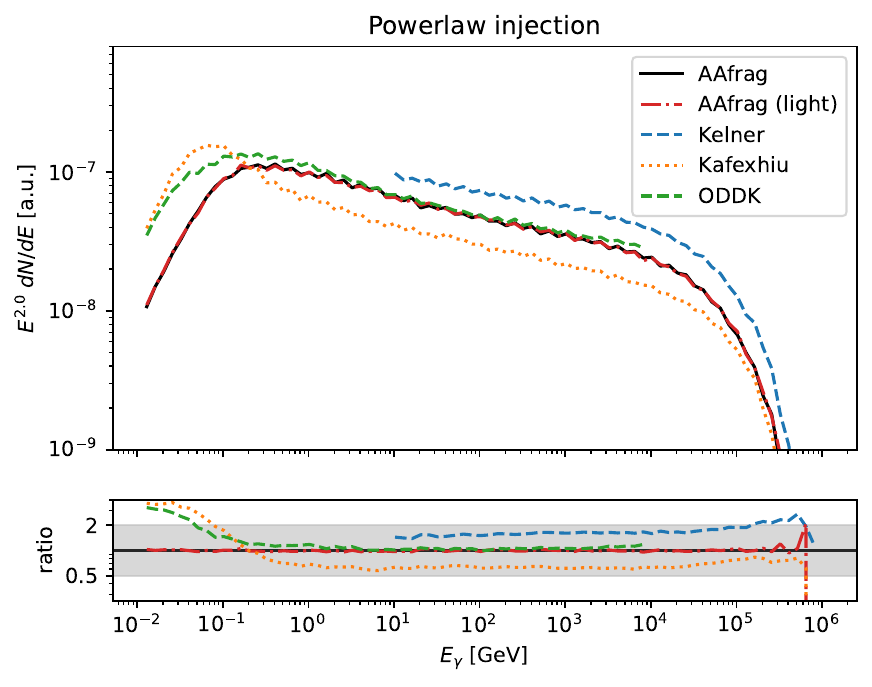} 
    \caption{\change{Same as Fig.\ \ref{fig:gamma_flux} but for the power-law injection $\diff N / \diff E \big|_\mathrm{source} \propto E^{-2.2}$.}}
    \label{fig:gamma_PL}
\end{figure*}

The resulting gamma-ray flux \change{assuming an injection with the local interstellar spectrum of CRs} for the different interaction models is shown in the upper panel of Fig.\ \ref{fig:gamma_flux} \change{and for the power-law injection in Fig.\ \ref{fig:gamma_PL}. Both assumptions for the source injection show the same trend and are discussed together in the following}. 

In the intermediate energy range $20 \, \mathrm{GeV} \leq E_\gamma \leq 5\cdot 10^{4} \, \mathrm{GeV}$ all models show the same energy scaling but differ in the overall normalization. At the highest energies, this trend continues except for the ODDK model, which is not shown due to the lack of cross-section data for $E_\gamma \geq 10^4 \, \mathrm{GeV}$. Also, the cut-off at $\sim 10^{6}$ GeV, based on the maximal proton energy ($E_\mathrm{p, max} = 10^{7}$ GeV) is visible in all models. 
In the lower energies, a clear difference between the cross-section models can be seen. The model by \cite{Kelner2006} does not cover primary energies $E_p < 100$ GeV and is therefore not shown for $E_\gamma < 10$ GeV, assuming that $\sim 10 \%$ of the energy goes into gamma-ray production. The models by \cite{Kafexhiu2014} and \cite{ODDK23} show an enhancement at the lowest energies compared to the AAfrag model.

The impact of multiple interactions of the protons is negligible in this GMC. This can be seen in the difference between the \textit{AAfrag} (black solid line) and \textit{AAfrag (light)} (red dash-dotted line) models. The model labeled "light" denotes only electrons, positrons, gamma rays, and neutrinos as secondaries, while the full model also allows for secondary protons. As the resulting gamma-ray flux stays the same in both models the impact of the up-scattered protons can be neglected.

In the lower panels of Fig.\  \ref{fig:gamma_flux} \change{and \ref{fig:gamma_PL}} the ratio between the different cross-section models is shown. Here, the \textit{AAfrag} model is used as a baseline. The ratio stays constant for most energy ranges as the spectral slope is mainly determined by the SED of the primary protons. But the normalization shows differences. In most cases, the uncertainty is within a factor $\sim 2$. Only in the sub-GeV regime, larger differences can be seen.

%\color{blue}
\section{Test application to a simplified Galactic plane} \label{sec:application_neutrino}

\change{To study the uncertainties in neutrino production, the new CRPropa module is applied to the Galactic plane. Here, we do not aim to derive a realistic description of the diffuse neutrino emission from the Milky Way. Therefore, all astrophysical inputs are kept as simple as possible. }

\subsection{Simulation setup}
\change{We solve the CR transport equation 
\begin{equation}
    \frac{\partial n}{\partial t} = 0 = \nabla(D\nabla n) - \frac{\partial}{\partial E} \left[ \frac{\diff E}{\diff t} n\right] + S(\vec{r}, E)
\end{equation}
for the differential CR number density $n = n(\vec{r}, E)$ in the steady state using stochastic differential equations implemented in the \texttt{DiffusionSDE} module of CRPropa \cite{Merten:2017mgk}. For the diffusion coefficient $D = D(E)$ the default parameters are used and the anisotropy in the diffusion is neglected. For the source term $S(\vec{r}, E)$ an energy scaling $S \propto E^{-2.2}$ is used\footnote{The CPRopa simulation itself uses an $E^{-1}$ distribution which is afterwards reweighted.}. The spatial source distribution follows the Galactic SNR distribution from \cite{Case96}. }

\change{The energy loss $\diff E/ \diff t$ is calculated using the hadronic interaction plugin presented in Sec.\ \ref{sec:crpropa}. 
This uses the analytical galactic density distribution from Nakanishi \etal \cite{Nakanishi2003, Nakanishi2006}, which is available in the CRPropa framework.
}

\change{The outer boundary of the Milky Way is assumed to be cylindrical with a radius of $R = 20$ kpc and a halo height of $H = 4$ kpc. Any pseudo-particles that reach the boundary are lost to the inter galactic medium. All neutrinos are directly written to the output and no further propagation is considered.
We ran the simulation for the Kelner \etal \cite{Kelner2006} and AAfrag \cite{AAfrag} cross section models, as they are the only ones that provide a description for neutrinos. A total of $N = 10^8$ pseudoparticles are simulated in each simulation. 
}

\begin{table}[tb]
    \centering
    \begin{tabular}{|c|c|}
        \hline
         module & parameters \\ \hline
         \multicolumn{2}{|c|}{\textbf{propagation}} \\ \hline
         \texttt{DiffusionSDE} & uniform magnetic field $\vec{b} = (0, 0, 1)$ \\ 
         & $l_\mathrm{min} = 1 \, \mathrm{kpc}\, , \, l_\mathrm{max} = 100 \, \mathrm{kpc}$ \\
         & anisotropy $\epsilon = 1$ (isotropic) \\ \hline
         \multicolumn{2}{|c|}{\textbf{interaction}} \\ \hline
         \texttt{HadronicInteraction} & cross-section model \\
         & density profile: Nakanishi \\ \hline 
         \multicolumn{2}{|c|}{\textbf{observer \& output}} \\ \hline
         \texttt{TextOutput} & \texttt{Event3D} \\
         \texttt{ObserverDetectAll} & \\
         \texttt{ObserverNucleusVeto} & \\ \hline 
         \multicolumn{2}{|c|}{\textbf{source}} \\ \hline 
         \texttt{SourceParticleType} & $(A, Z) = 1,1$ \\
         \texttt{SourcePowerLawSpectrum} & $E_\mathrm{min} = 1 \, \mathrm{GeV}\, , \, E_\mathrm{max} = 1 \, \mathrm{PeV}\, , \, \alpha = -1$ \\
         \texttt{SourceSNRDistribution} & \\
         \texttt{SourceIsotropicEmission} & \\ \hline 
         \multicolumn{2}{|c|}{\textbf{boundary}}    \\ \hline
         \texttt{MaximumTrajectoryLength} & $D_\mathrm{max} = 10 \, \mathrm{Gpc}$ \\
         \texttt{CylindricalBoundary} & $\vec{o} = (0, 0, - 4\, \mathrm{kpc})\, , \, H = 8\, \mathrm{kpc}\, , \, R = 20 \, \mathrm{kpc}$ \\ \hline
    \end{tabular}
    \caption{\change{CRPropa Modules used for the simulation of the Galactic neutrino production.}}
    \label{tab:modules_neutrinos}
\end{table}

\subsection{Result}
\begin{figure}[t]
    \centering
    \includegraphics[width=\linewidth]{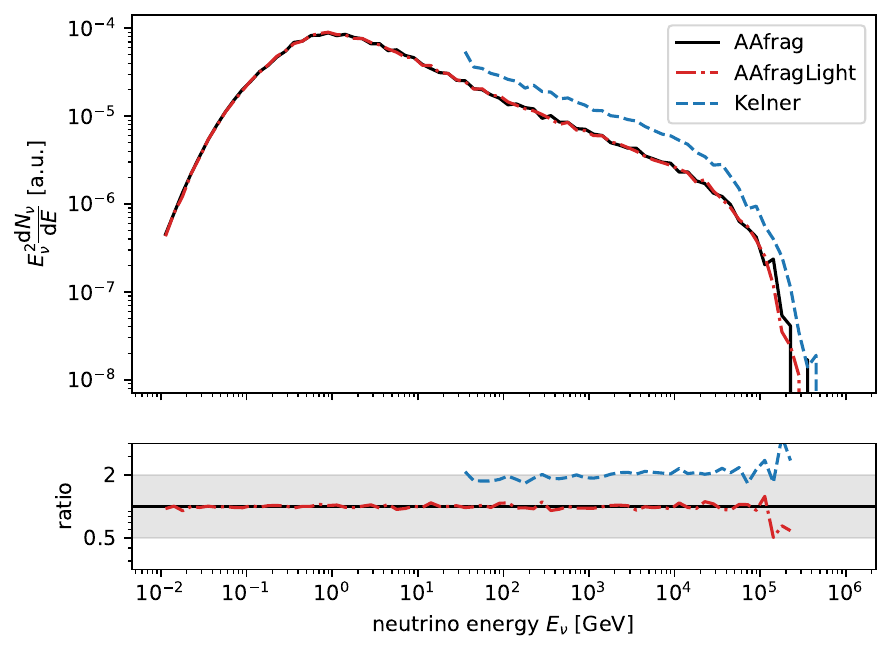}
    \caption{\change{\textit{Upper:} All-flavor neutrino flux from the simplified Galactic plane based on different cross section models. \textit{Lower:} ratio between the predictions. Here, the AAfrag model is used as a baseline. The gray band indicates a difference of a factor 2.}}
    \label{fig:neutrino_flux}
\end{figure}

\change{The resulting all-flavor neutrino flux is calculated using 80 energy bins logarithmically spaced  between $10^{-2}$ GeV and $10^6$ GeV. Fig.\ \ref{fig:neutrino_flux} shows the flux (upper panel) and the ratio to the AAfrag prediction (lower panel). }

\change{The small differences between the full AAfrag\cite{AAfrag} and the "AAfragLight" model, which neglects secondary protons and neutrons, can be explained by statistical fluctuations. The overall agreement between these two predictions suggests that the emission of neutrinos from the nuclear decay of neutrons, as well as the second interaction of the secondary protons, can be neglected. }

\change{The prediction from the Kelner \etal model \cite{Kelner2006} follows the same energy scaling as the AAfrag model, but the normalization is a factor $\sim$ 2 higher. As the model is derived for CR energies $E_p > 100$ GeV, we only show neutrinos with $E_\nu > 30$ GeV, assuming that $\sim 1/3$ of the primary energy goes into neutrino production. At the highest energies ($E > 10^5$ GeV), the statistical fluctuation becomes larger, leading to more differences between the cross section models.}
\section{Conclusions and outlook \label{conclusions:sec}}
In this work, we investigate the uncertainties in astrophysical gamma-ray \change{and neutrino} fluxes arising from different proton-proton cross-section models in the GeV to PeV energy range. We implemented four independently parametrized hadronic interaction models into the publicly available transport code CRPropa. We applied these models to a generic setup of a nearby giant molecular cloud to quantify the impact of cross-section choices on the resulting gamma-ray flux. \change{The effect on the neutrino flux is demonstrated by applying it to a simplified Galactic plane.}

Our results demonstrate that the choice of cross-section model can lead to significant variations in the predicted flux, with differences up to a factor of $\sim 2$. These differences are present over the full energy range that is relevant to produce a multimessenger prediction for Galactic, but also extragalactic cosmic-ray sources. \change{For the gamma-ray flux} largest deviations are found between the model of Kelner \cite{Kelner2006} and Kafexhiu \cite{Kafexhiu2014}, with Kelner resulting in a flux that is systematically larger by a factor of 3. Kafexhiu et al. \cite{Kafexhiu2014} discuss that $\pi^0$-production with QGSJet-I, as used by \cite{Kelner2006} is generally a factor of ~1.7 larger than QGSJet-II used by \cite{Kafexhiu2014}, which would explain part of the systematic difference. When comparing the model by Kafexhiu with AAfrag \cite{AAfrag}, the latter is systematically larger by a factor of close to 2 above 1 GeV, but lower below. AAfrag and ODDK \cite{ODDK22, ODDK23} are similar above $\sim 0.3$~GeV, and ODDK gives a significantly larger flux below 0.3~GeV. 
\change{The differences in the neutrino flux prediction follow the same behavior and the Kelner \cite{Kelner2006} model predicts a flux $\sim 2$ times higher flux than that of AAfrag \cite{AAfrag}.}
It should be mentioned that models based on LHC tuned data are not recommended to be used below 100 GeV, as they are tuned above this value. The processes described and the ad-hoc hadronization procedures are not generally applicable below the tuning range. In the case of QGSJet-based models, no distinction between the soft and hard processes are being made (unlike EPOS or Sybill) and the systematic error in applying it toward lower energies thus can also be lower, although still hard to quantify.

Our results highlight the importance of accurate cross-section modeling in interpreting multimessenger data and identifying cosmic-ray sources. In this context, the implementation of tables from the codes of hadronic interactions, as mentioned above, is very valuable, in particular by using frameworks like \texttt{crmc}~\citep{ulrich_2021_5270381} and  \texttt{chromo}~\citep{chromo_repository} which allow sampling the hadronic generators within a common interface and create production tables for secondaries to be used for direct sampling or to fit analytic expressions describing the distributions used for sampling.

The implementation of such a systematic approach to treat hadronic interactions quantitatively in CRPropa will be the next step toward a better understanding of multimessenger sources. This first step of having different descriptions at hand to quantify uncertainties, and the next to use the interaction models directly provides a valuable tool for the astrophysics community to study the propagation and interaction of cosmic rays in various astrophysical environments. The modular structure of CRPropa allows for easy integration of new interaction models and the extension of the code to include additional physical processes.

\acknowledgments
We acknowledge support from the Deutsche Forschungsgemeinschaft DFG, via the Collaborative Research Center SFB1491 "Cosmic Interacting Matters - From Source to Signal" (project No. 445052434), and via \textit{MICRO} (project number 445990517).

\appendix
\section{Testing of yields} \label{app:yields}
In this appendix, a more detailed test for the yields of the secondaries is presented. In section \ref{ssec:test_yields} the test was already performed for the cross-section model \textit{AAfrag} \cite{AAfrag}. Here we extend the test to all models presented in section \ref{ssec:diff_cross}.

The test is designed to probe the yields of the produced secondaries. Here, we \change{show the secondary $\gamma$-rays, electrons ($e^-$), positrons ($e^+$) and neutrinos ($\nu_e$, $\nu_\mu$) where they are included.}

\begin{figure*}[t]
    \centering
    \includegraphics[width=\textwidth]{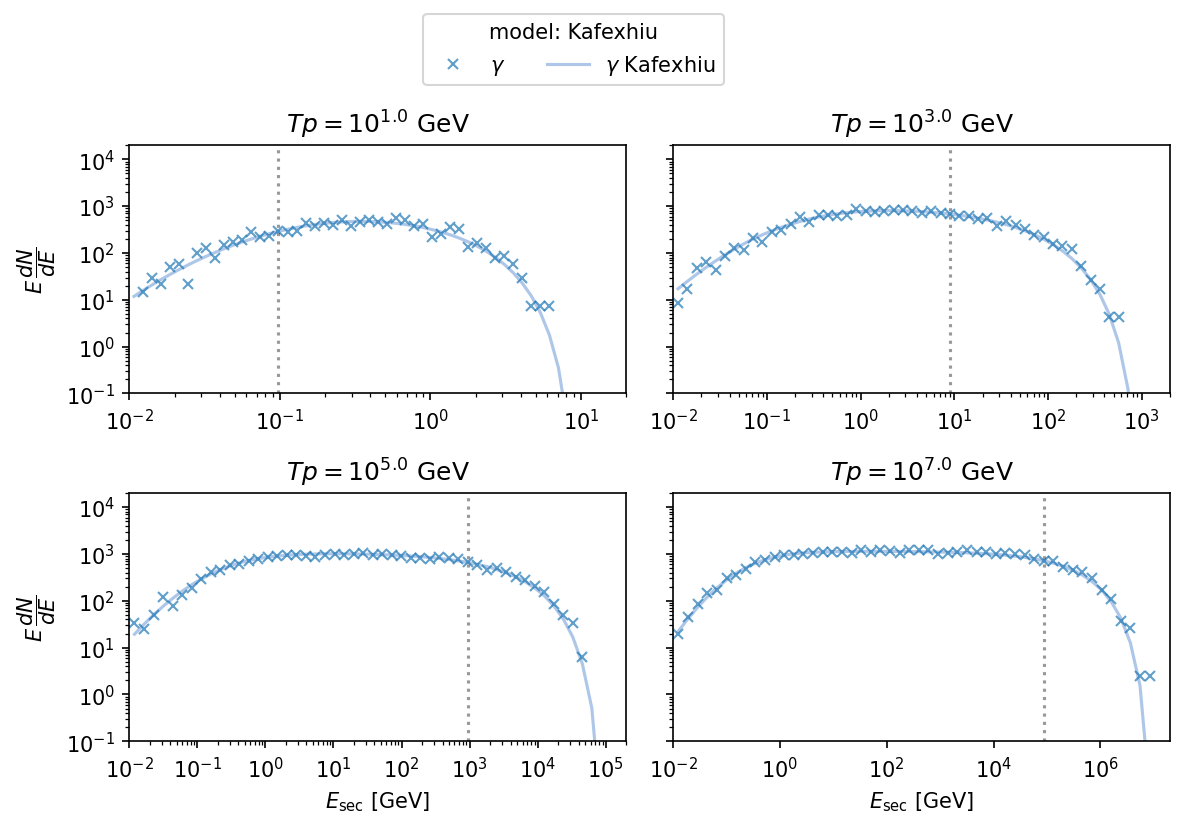}
    \caption{Secondary yields derived from the cross-section model from \cite{Kafexhiu2014}. For comparison the shape of the differential cross-section, normalized at $E_\mathrm{sec} = 0.01\, T_p$ is shown.}
    \label{app:fig:yields_Kafexhiu}
\end{figure*}

The spectrum for the cross-section model by \cite{Kafexhiu2014} is shown in Fig.~\ref{app:fig:yields_Kafexhiu}. It only contains the secondary gamma rays as described before. The sampled data are in good agreement with the prediction from the differential cross-section. 

\begin{figure*}[tb]
    \centering
    \includegraphics[width=\textwidth]{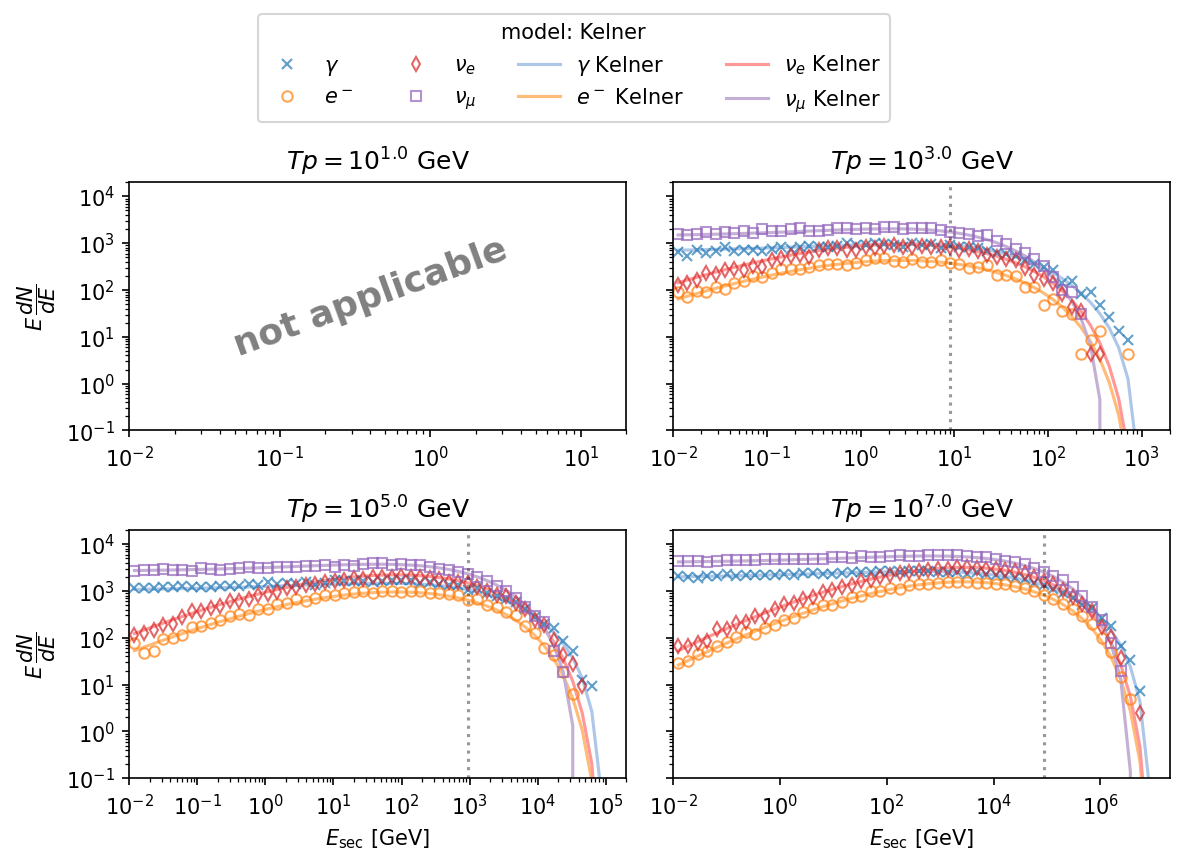}
    \caption{Same as Fig.~\ref{app:fig:yields_Kafexhiu} but for the cross-section model by Kelner \cite{Kelner2006}. Note that this model is only valid for $T_p > 100$ GeV.}
    \label{app:fig:yields_Kelner}
\end{figure*}

The spectrum for the cross-section model from Kelner \etal \citep{Kelner2006} is shown in Fig.~\ref{app:fig:yields_Kelner}. This model is not applicable for kinetic energies $T_p < 100$ GeV. The sampled secondaries are in good agreement with the prediction. Kelner \etal do not discriminate between electrons and positrons. Therefore, \change{only the sampled electrons are shown.} 

\begin{figure*}[tb]
    \centering
    \includegraphics[width=\textwidth]{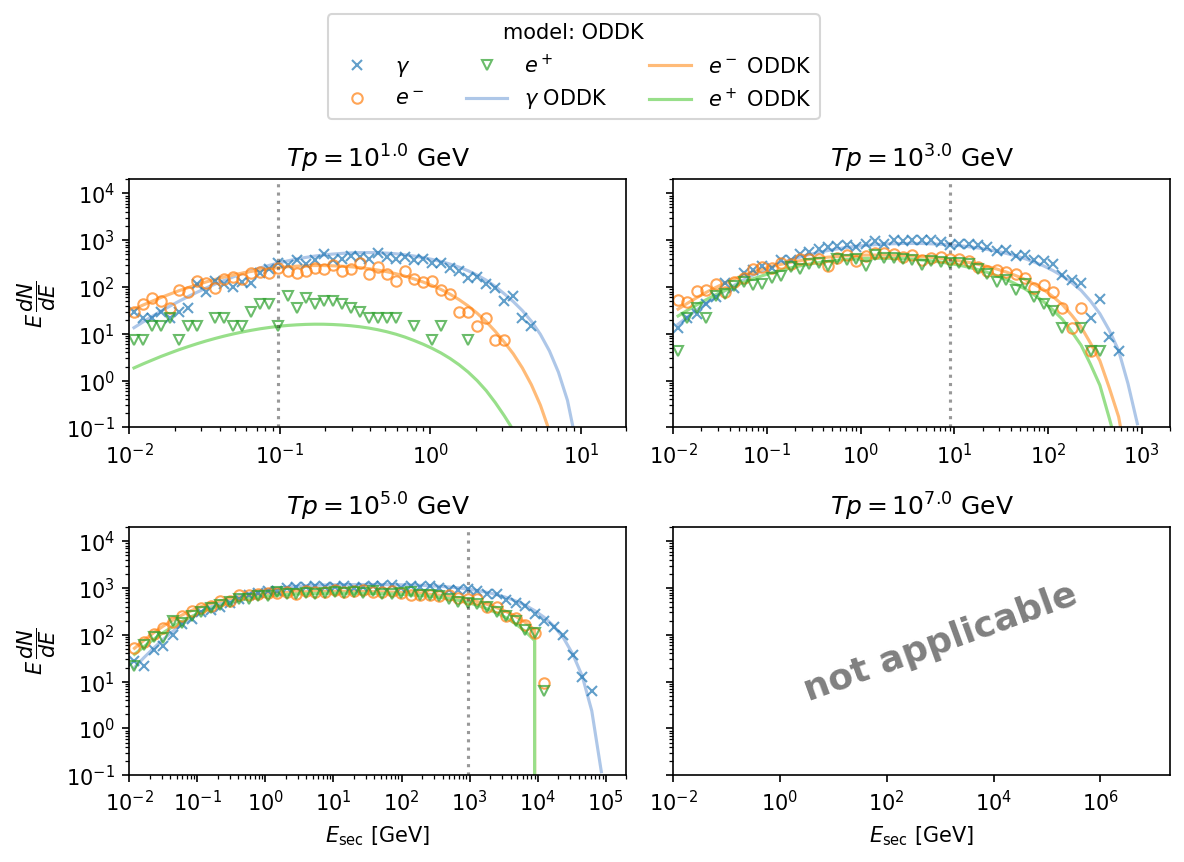}
    \caption{Same as Fig.~\ref{app:fig:yields_Kafexhiu} but for the cross-section model ODDK \citep{ODDK22, ODDK23}. Note that this model is only applicable for the lepton production at $T_p < 10^{4}$ GeV and the gamma-ray production at $T_p < 10^{5}$ GeV.}
    \label{app:fig:yields_ODDK}
\end{figure*}

The spectrum for the cross-section model ODDK \citep{ODDK22, ODDK23} is shown in Fig.~\ref{app:fig:yields_ODDK}. The sampled spectra in the lower energies agree well. Only at the highest energies the limit of the tabulated cross-sections are visible. Therefore this model should only be used up to a primary energy of $T_p = 10^4 \, \mathrm{GeV}$ or combined with the description of another cross-section model.

\clearpage

% Bibliography

%% [A] Recommended: using JHEP.bst file
\bibliographystyle{JHEP}
\bibliography{bib.bib}

\end{document}